\documentclass[10pt]{ismd08}
\usepackage{graphicx}
\usepackage{cite,./mcite}

\begin{document}
\title{Progress in Parton Distribution Functions \\
and Implications for LHC\footnote{~~Presented at the XXXVIII International Symposium on Multiparticle Dynamics (ISMD 2008), 15--20 September 2008, DESY, Hamburg, Germany.}}
\author{W. James Stirling}
\institute{Department of Physics, University of Cambridge, Cambridge CB3 0HE, UK}
\begin{flushright}
CAVENDISH-HEP-08-15 \\
December 2008
\end{flushright}
\maketitle
\begin{abstract}
Parton distribution functions (pdfs) are an important ingredient for LHC phenomenology. Recent progress in determining pdfs from global analyses is reviewed, and some of the most important outstanding issues are highlighted. Particular attention is paid to the precision with which predictions for LHC `standard-candle' cross sections can be made, and also to new information that LHC can provide on pdfs.
\end{abstract}

\def\beq{\begin{equation}} 
\def\eeq{\end{equation}} 
\def\beqn{\begin{eqnarray}} 
\def\eeqn{\end{eqnarray}} 
\def\as{\alpha_S} 
\def\asmz{\alpha_S(M_Z^2)} 
\def\sighat{\hat{\sigma}} 

\section{Introduction}

High-precision cross-section predictions for both Standard Model and Beyond Standard Model processes at the LHC require high-precision parton distribution functions (pdfs). In some cases, the uncertainty in our knowledge of the pdfs is a significant or even dominant part of the overall uncertainty in the theoretical prediction. Of course, the more accurate the signal and background predictions, the easier it will be to detect new physics. Fortunately the LHC provides a number of `standard-candle' processes, whose measured cross sections can be used to check the theoretical framework (factorisation, DGLAP evolution etc.). The paradigms are $\sigma(Z)$ and $\sigma(W)$, for which there are realistic prospects of experimental measurements and theoretical predictions accurate at the few \% level.

At the same time, the LHC can provide new information on the pdfs themselves. Hadron collider data have always been an important ingredient of pdf global fits. For example, fixed-target Drell-Yan data currently provide (unique) information on high-$x$ sea quarks, Tevatron high-$E_T$ jet data provide direct information on the high-$x$ gluon, and Tevatron $W$ and $Z$ cross sections and distributions provide information on quark distributions complementary to that from deep inelastic scattering. There is every prospect that similar measurements at the LHC will improve our knowledge of pdfs even further. 

The basic theoretical tool for precision predictions for hadron colliders such as the Tevatron and the LHC is the QCD factorization theorem for short-distance inclusive processes: 
\begin{equation}
\sigma_{AB}\>  =\>  \int dx_a dx_b\>  f_{a/A}(x_a,\mu_F^2) \>
f_{b/B}(x_b,\mu_F^2)\;
\times \;  [\> \sighat_0\> +\> \as(\mu_R^2)\> \sighat_1\> +\> ...\> ]_{ab\to X} \; .
\label{signlo}
\end{equation}
Formally, the cross section calculated to all orders in perturbation theory is invariant under changes in the factorization scale ($\mu_F$) and the renormalization scale ($\mu_R$), the scale dependence of the coefficients  $\sighat_0, \sighat_1, ...$
exactly compensating the explicit scale dependence of the pdfs and the QCD coupling constant.
This compensation becomes
more exact as more terms are included in the perturbation series. In the absence of a complete set of higher-order corrections, it is necessary to make a specific choice for the two scales in order to make cross-section predictions. A variation of the scales by a factor of 2 around some `natural' scale $M$ for the process, i.e. $M/2 < \mu_F, \mu_R < 2M$, is often used\footnote{~Care should be taken when comparing scale uncertainties produced in this way. Some authors set $\mu =  \mu_F = \mu_R  $  and vary $\mu$ in the standard range, while others either allow $\mu_F$  and $ \mu_R $  to vary independently in the range, or place additional restrictions on $r = \mu_F / \mu_R $, e.g. $1/2 < r < 2$.} to characterise the uncertainty from unknown higher-order terms in the series. The overall theoretical error on a cross section prediction can then be estimated as $\delta\sigma_{\rm th}^2 = \delta\sigma_{\rm pdf}^2 + \delta\sigma_{\rm scl}^2$.

Almost all the theoretical quantities (subprocess cross sections, coefficient functions and splitting functions) that are needed for a global fit are nowadays known to NNLO in pQCD, and so this will be the {\it de facto} benchmark for LHC phenomenology. In some cases, e.g. $W$ and $Z$ production, electroweak corrections are also known and can be included. The following table illustrates the relative size of the pdf and scale uncertainties for some standard processes at the LHC, calculated at NNLO\footnote{In the case of $t \bar t$ production, an approximation to the (as yet uncalculated) full correction has been derived, see \cite{Moch:2008qy}.} in pQCD.
Here the pdf uncertainties are taken from the recent MSTW global fit \cite{graeme_dis08,MSTW2008}, while the scale uncertainty estimates for $t\bar t$ and Higgs production are taken from Refs.~\cite{Moch:2008qy} and \cite{Harlander:2002wh} respectively. Evidently the pdf uncertainty is a significant issue for $Z$ and $t\bar t$ production, but not at present for Higgs production.
\begin{center}
\begin{tabular}{|l|c|c|}
\hline
 process & $\quad\delta\sigma_{\rm pdf}\quad$ & $\quad\delta\sigma_{\rm scl}\quad$ \\
\hline
$pp\to Z + X$ & $\pm 2\%$ & $\pm 1\%$ \\
$pp\to t\bar t + X$ & $\pm 2\%$ & $\pm 3\%$ \\
$pp\to H (120~{\rm GeV}) + X$ & $\pm 2\%$ & $\pm 15\%$
\\
\hline
\end{tabular}
\end{center}

\section{How pdfs are obtained}

\begin{figure}[th]
%\vspace{2.0cm}                                                              
\hspace{2.0cm}  
\begin{centering}\includegraphics[width=0.5\columnwidth,keepaspectratio,angle=270]{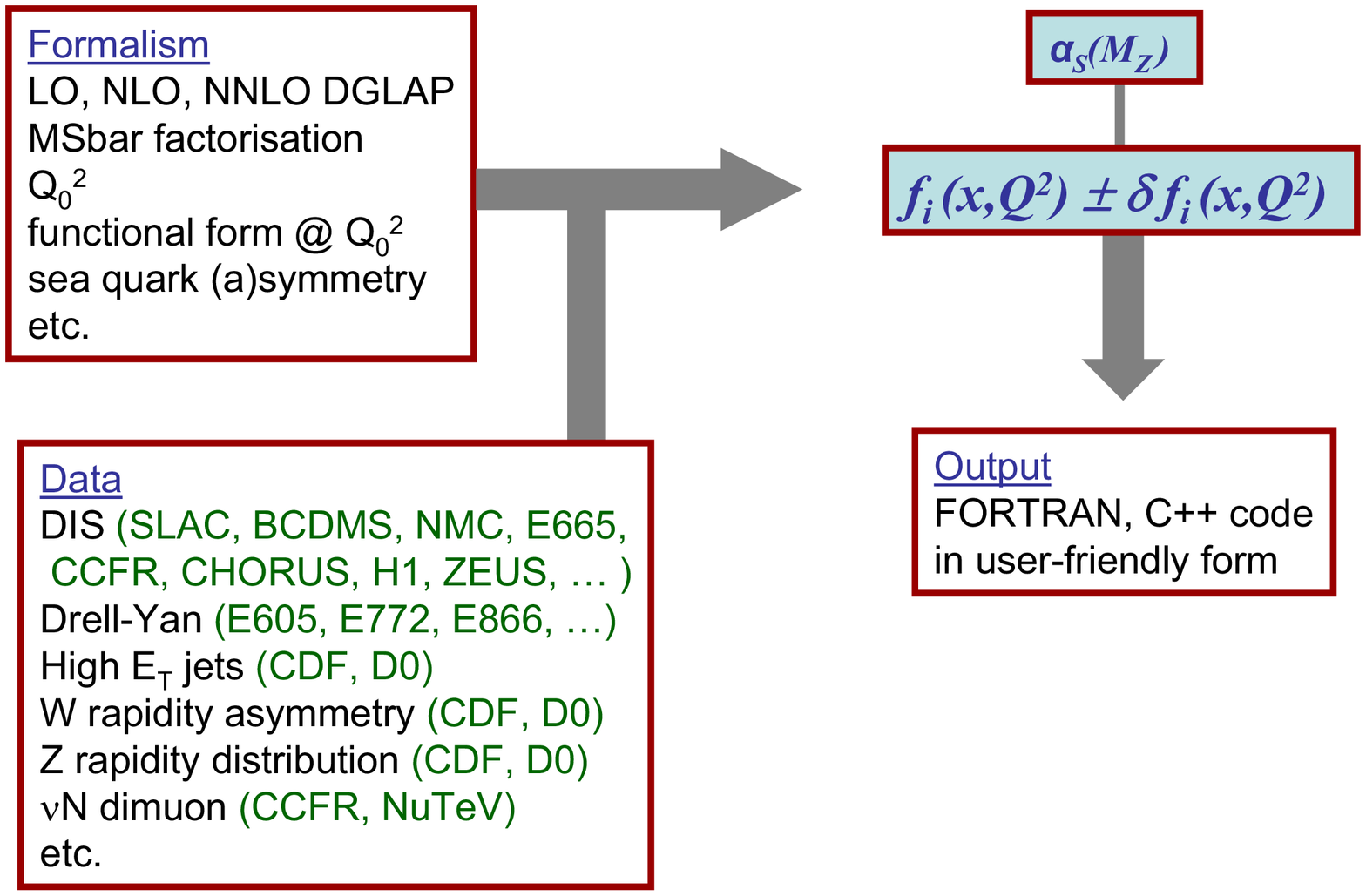}
\end{centering}                                                          
\vspace{0.5cm}
\caption{Anatomy of a pdf global fit.\label{fig:pdfmethod}}                    
\end{figure}

The method by which pdfs are obtained from a global fit to a variety of `hard scattering' data is by now well known -- a schematic summary is shown in Fig.~\ref{fig:pdfmethod}. A typical set of input data (as used, for example, by the MSTW and CTEQ collaborations, see Section~\ref{sec:progress}) is given in the following Table, together with the partons that they constrain.
\bigskip

  \begin{tabular}{|l|l|}
    \hline
H1, ZEUS & {$F^{e^{+}p}_2(x,Q^2)$, 
$F^{e^{-}p}_2(x,Q^2)$}\quad NC + CC
\\
BCDMS & {$F^{\mu p}_2(x,Q^2), F^{\mu d}_2(x,Q^2)$}
\\
NMC & {$F^{\mu p}_2(x,Q^2), F^{\mu d}_2(x,Q^2),
F^{\mu n}_2(x,Q^2)/F^{\mu p}_2(x,Q^2)$}
\\
SLAC & {$F^{e^- p}_2(x,Q^2), F^{e^- d}_2(x,Q^2)$}
\\
E665 & {$F^{\mu p}_2(x,Q^2),F^{\mu d}_2(x,Q^2)$}
\\
CCFR, NuTeV, CHORUS & {$F^{\nu(\bar\nu) N}_2(x,Q^2),
 F^{\nu(\bar\nu) N}_3(x,Q^2)$}
\\
& ~~~~$\to$ { $q$, $\bar q$} at all $x$ and $g$ at medium, small $x$
\\
H1, ZEUS & {$F^{e^{\pm}p}_{2,c}(x,Q^2), F^{e^{\pm}p}_{2,b}(x,Q^2)$}  $\to$ { $c,b$}
\\
E605, E772, E866 & Drell-Yan {$ p N \to \mu \bar \mu + X$} $\to$
{ $\bar q$ ($g$)}
\\
E866 & Drell-Yan $p,n$ asymmetry $\to $ { $\bar u, \bar d$}
\\
CDF, D0 &  $W^\pm$  rapidity asymmetry $\to $ { $u/d$} ratio at high $x$
\\
CDF, D0 & $Z^0$  rapidity distribution $\to $ { $u, d$}
\\
CDF, D0 & inclusive jet data $\to $  $g$ at high $x$
\\
H1, ZEUS & DIS $+$ jet data $\to $ { $g$} at medium $x$
\\
CCFR, NuTeV  & dimuon data
$\to $ strange sea { $s$, $\bar s$}
\\
    \hline
  \end{tabular}
%\vspace{1cm}
\bigskip

 Over the past 15 years, the quality and quantity of the data has improved enormously, so that nowadays the pdfs are known to very high accuracy, typically to within a few $\%$ over a broad range of $x$ away from $x=0,1$. In terms of recent developments, much attention has been focused on the  heavy quark ($s,c,b$) distributions.
Until recently, the strange quark distribution was generally parametrised as
\beq
s(x,Q_0^2) = \bar s(x,Q_0^2) = \frac{\kappa}{2} \left[  \bar u(x,Q_0^2) + \bar d(x,Q_0^2)\right] 
\eeq
with $\kappa = 0.4 - 0.5$ suggested by (neutrino DIS) data. The suppression was understood as a non-perturbative mass effect. Recent measurements of dimuon production in $\nu N$ DIS (for example, by CCFR and NuTeV) allow a more-or-less direct determination of both $s$ and $\bar s$, via
\beq
\frac{d\sigma}{dxdy}\left (\nu_\mu(\bar\nu_\mu) N \to \mu^+\mu^-X\right) = B_c {\cal N A}  
\frac{d\sigma}{dxdy}\left (\nu_\mu s(\bar\nu_\mu\bar s) \to c\mu^-(\bar c \mu^+)\right)
\eeq
in the range $0.01 < x < 0.4$. The data appear to slightly prefer $s(x,Q_0^2) \neq \bar s(x,Q_0^2)$, both having a different shape to the light sea quark distributions. Generalised parametrisations for $s$ and $\bar s$ are therefore used in the most recent global fits. 

The charm and bottom quarks are considered sufficiently massive to allow a pQCD treatment, i.e. the distributions are assumed to be generated perturbatively via $g \to Q \bar Q$. Two regimes can be distinguished: (i) $Q^2 \sim m_Q^2$ where it is essential to include the {\it full} $m_Q$ dependence in order to get the correct threshold behaviour, and (ii) $Q^2 \gg m_Q^2$ where the heavy quarks can be treated as essentially massless partons, with large logarithmic contributions of the form $\alpha_S^n \ln^n(Q^2/m_Q^2)$ automatically resummed by the DGLAP equations. The so-called Fixed Flavour Number Scheme (FFNS), in which heavy quarks are not treated as partons, is only valid in region (i), whereas the Zero Mass Variable Flavour Number Scheme (ZM-VFNS), in which heavy quarks evolve as massless partons from zero at threshold, applies to region (ii) only. In recent years, a more general set of General Mass Variable Flavour Number Schemes (GM-VFNS) have been developed, with the advantage of interpolating smoothly and consistently between the two $Q^2$ regions, at a given order (up to and including NNLO in practice) in perturbation theory.  The two most important points to note are: (i) the definition of a consistent GM-VFNS is tricky and non-unique (not least due to the assignment of ${\cal O}(m_Q^2/Q^2)$ contributions), and implementation of an improved treatment of heavy quarks can have a significant knock-on effect on light partons, and (ii) GM-VFNS predictions for the structure functions $F_2^{c\bar c} $ and $F_2^{b \bar b}$ agree well with measurements at HERA. A more detailed discussion of the treatment of heavy quark pdfs can be found in \cite{Thorne:2008xf}.

Another major advance in recent years has been the treatment of uncertainties in the distribution functions, and most global fit groups produce `pdfs with errors' sets. These are of course useful in assessing the error on cross-section predictions due to the pdfs themselves. A typical package will consist of a `best fit' set and $\sim$ 30--40 error sets designed to reflect a $\pm 1 \sigma$ variation of all the parameters used to define the starting distributions, as determined by the uncertainties on the data used in the global fit. However, in addition to these `experimental' uncertainties, there are also uncertainties associated with theoretical assumptions and/or prejudices in the way the global fit is set up and performed. Although these are generally more difficult to quantify, they are often the main reason for the differences between the sets produced by different groups.
The following is a non-exhaustive list of the reasons why `best fit' pdfs and errors can differ:

\noindent$\bullet\ \ $ different data sets in the fit:\\
\indent--- different subselection of data \\
\indent--- different treatment of experimental systematic errors\\
\noindent$\bullet\ \ $ different choice of:\\ 
\indent--- pQCD order (in DGLAP and cross sections)\\
\indent--- factorisation/renormalisation scheme/scale\\
\indent--- $Q_0^2$\\
\indent--- parametric form $f_i(x,Q_0^2) = Ax^a(1-x)^b c(x)$ etc., and implicit extrapolation\\
\indent--- $\alpha_S$\\
\indent--- treatment of heavy flavours\\
\indent--- theoretical assumptions about $x \to 0,1$ behaviour\\
\indent--- theoretical assumptions about sea quark flavour asymmetry\\
\indent--- $\chi^2$ tolerance to define $\pm \delta f_i$\\
\indent--- evolution, cross-section codes, rounding errors etc.

\noindent Note that these can apply both to comparisons of the type CTEQ {\it vs.} MRST {\it vs.} ... and to CTEQ6.1 {\it vs.} CTEQ 6.5 etc.

\section{Recent progress}
\label{sec:progress}

There are a number of groups producing pdf sets from global fits to data. In this section we give a very brief summary of these, with references to where more information can be found.

The Martin--Stirling--Thorne--Watt {\bf MSTW} (formerly Martin--Roberts--Stirling--Thorne {\bf MRST}) collaboration produces sets at LO, NLO and NNLO using a `maximal' set of fitted data as described in the previous section. The previous MRST2006 sets \cite{Martin:2007bv} contained an update of the NNLO fit to include both pdf errors and an improved GM-VFNS treatment of $c$ and $b$. The new MSTW2008 sets \cite{graeme_dis08,MSTW2008} include (i) new data sets in the fit (CHORUS and NuTeV neutrino data and HERA DIS$+$jet data), (ii) a more sophisticated treatment of $s$ and $\bar s$ in which both are allowed to have independent shapes and normalisations, and (iii) an improved treatment of the tolerance procedure to define the error sets (for a summary see \cite{graeme_dis08}).

The {\bf CTEQ} collaboration (Ref.~\cite{Nadolsky:2008zw} and references therein) produces LO and NLO pdf sets from global fits using roughly the same maximal data set as MSTW/MRST. Earlier this year, the previous (2006) 6.5 set was updated to produce set 6.6. CTEQ 6.5 was characterised by the first implementation of a GM-VFNS (the `SACOT$-\chi$' scheme \cite{Kretzer:2003it,Tung:2001mv}), which had a significant impact on the $c$ and $b$ distributions, a compensating impact on the $u$ and $d$ partons, and a corresponding change in the predictions for $\sigma(W,Z)$. The new 6.6 set has a more sophisticated treatment of the $s$ and $\bar s$ pdfs, allowing these to have a more general shape and normalisation than previously. The impact of an additional `intrinsic charm' contribution is also studied.

\begin{figure}[th]
\noindent 
\begin{centering}\includegraphics[width=0.5\columnwidth,keepaspectratio]{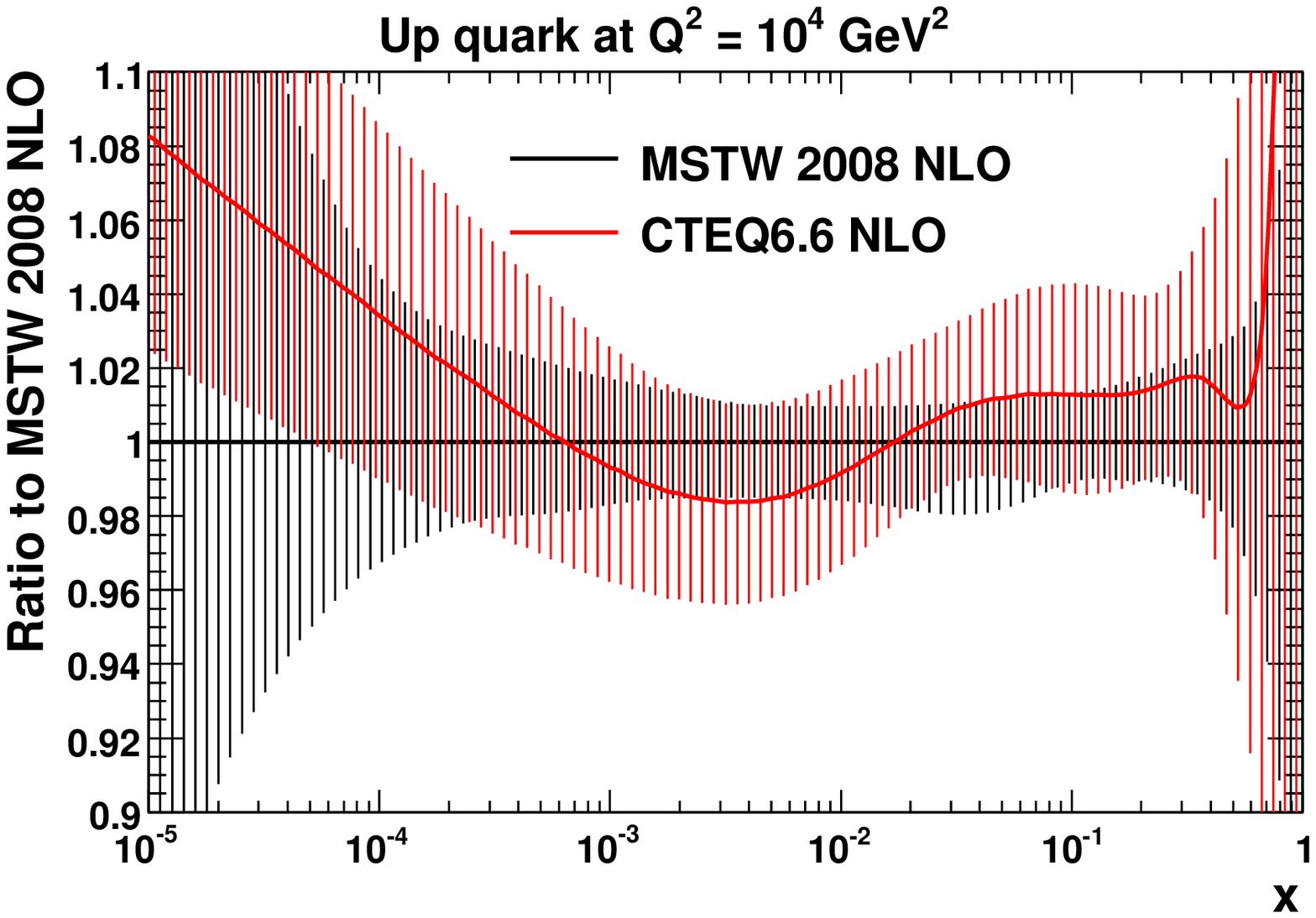}
\includegraphics[width=0.5\columnwidth,keepaspectratio]{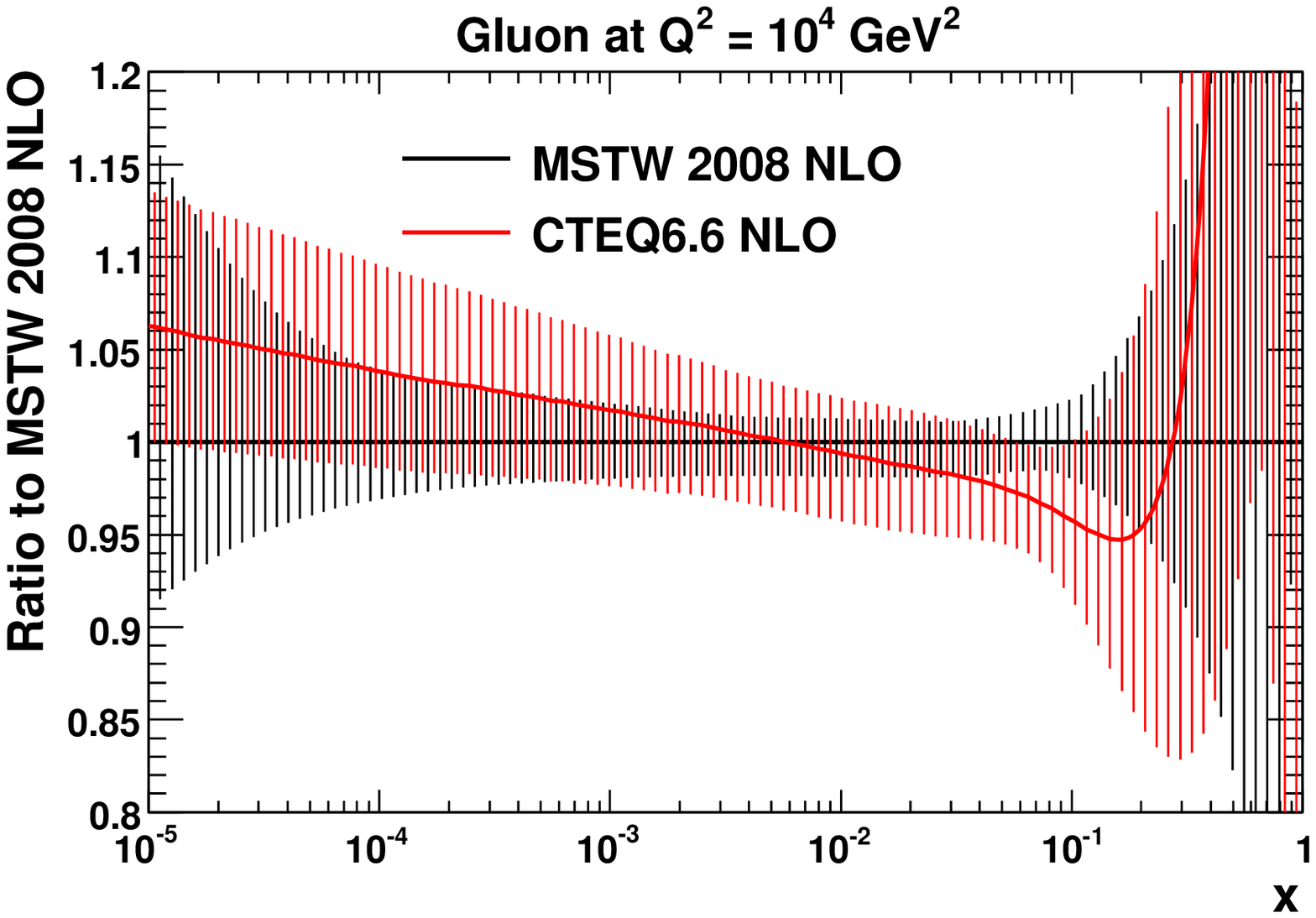}
\end{centering}
\vspace{-0.5cm}
\caption{Comparison of recent MSTW and CTEQ up quark (left) and gluon (right) NLO parton distributions.\label{fig:mstwcteq}}
\end{figure}

Given the similarity of the data fitted and the theoretical framework used, it is no surprise that the pdf outputs from the MSTW and CTEQ analyses are similar. This is illustrated Fig.~\ref{fig:mstwcteq}, which  compares the latest MSTW2008 and CTEQ6.6 NLO $u$ and $g$ pdfs (with errors) at $Q^2 = 10^4$~GeV$^2$. Note that the broader CTEQ error band is in part a reflection of a different choice of tolerance in defining the allowed range of $\Delta\chi^2$. The MSTW gluon is smaller at small $x$, because the parameterisation at $Q_0^2 = 1$~GeV$^2$ allows the starting distribution to be negative at small $x$, unlike in the CTEQ (central) fits where the gluon is always constrained to be positive.

{\bf Alekhin et al.} produce sets at LO, NLO and NNLO. The original 2002 (Alekhin) set \cite{Alekhin:2002fv} was updated first in 2006 
\cite{Alekhin:2006zm} (Alekhin--Melnikov--Petriello) and again in 2007 \cite{Alekhin:2008ua} (Alekhin--Kulagin--Petti). The 2002 fit was based on DIS structure function data only (SLAC, BCDMS, NMC, E665, H1, ZEUS). The 2006 AMP version added E605 and E866 Drell-Yan data, and CHORUS, CCFR and NuTeV neutrino structure function and dimuon data. Unlike the CTEQ and MSTW/MRST fits, the Alekhin fit does not include Tevatron high-$E_T$ jet data, nor a complete GM-VFNS treatment of heavy quarks, and this accounts for much of the differences between the resulting parton distributions.

Both the {\bf H1} and {\bf ZEUS} collaborations have produced pdf sets in the past based on their own HERA DIS data supplemented by other DIS data. The most recent H1 (2003) set added BCDMS data to H1 structure function data to give a broad coverage in $x$ and $Q^2$. The ZEUS (2005) set was based on ZEUS data (both inclusive structure function and DIS$+$jet  data)  only. The two collaborations also had different treatments of pdf errors: offset (ZEUS) {\it vs.} Hessian (H1). Recently H1 and ZEUS have joined together to produce a combined pdf set, HERAPDF0.1, details of which can be found in the talk by Gang Li \cite{H1ZEUS}. Differences between the previous H1 and ZEUS fitting procedures have been resolved, and experimental and model uncertainties have been carefully considered. However this fit uses only HERA inclusive cross section NC and CC $e^\pm p$ data, and therefore there are small but significant differences in both quark and gluon differences in comparison with MSTW and CTEQ, which can in large part be traced to the influence of Tevatron and fixed-target Drell-Yan data in the latter global fits. 

The {\bf NNPDF} (Neural Net) collaboration \cite{NNPDFtalk} uses neural net technology  in the fit to avoid having to choose a particular parametric form at $Q_0^2$. The new (NLO) set, NNPDF1.0, is based on a Monte Carlo approach, with neural networks used as unbiased interpolants. The method is designed to provide a faithful and statistically sound
representation of the uncertainty on parton distributions. The fit is performed to a restricted `DIS only' data set in a ZM-VFNS scheme for the heavy quarks. The absence of Drell-Yan and neutrino dimuon data from the fit means that the detailed flavour structure of the quark sea is not well determined (and therefore neither are the predictions for $W$ and $Z$ cross sections at the LHC, see Section~\ref{sec:LHCimpact} below). The absence of Tevatron High-$E_T$ jet data from the fit is another signficant source of difference between NNPDF and CTEQ/MSTW. A recent update (NNPDF1.1 \cite{NNPDFupdate}) introduces independent parametrisations for the strange pdfs. 

Finally, there have been a number of other studies of pdfs designed for particular purposes or to investigate different theoretical frameworks. For example, the `dynamical parton model' approach (see \cite{JimenezDelgado:2008hf} and references therein) attempts to describe DIS and other data from a set of valence-like partons evolved upwards in $Q^2$ from a low starting scale. A reasonable fit is obtained, although the total $\chi^2$ is signficantly larger than in a (standard) fit in which the small-$x$ parameters are unconstrained.

\begin{figure}[t]
\noindent 
\vspace{-1.0cm}
\begin{centering}\includegraphics[width=0.48\columnwidth,keepaspectratio]{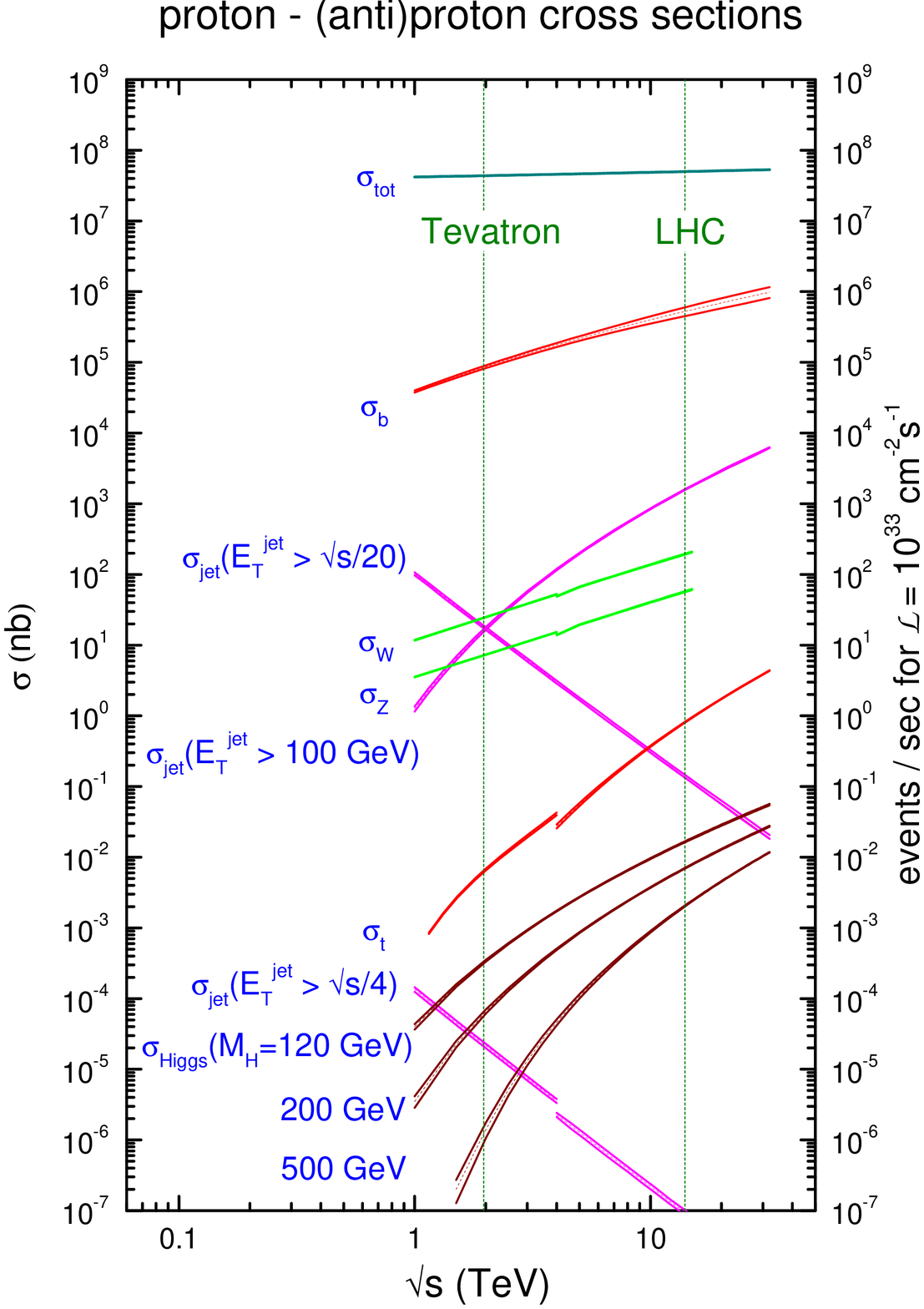}\includegraphics[width=0.52\columnwidth,keepaspectratio]{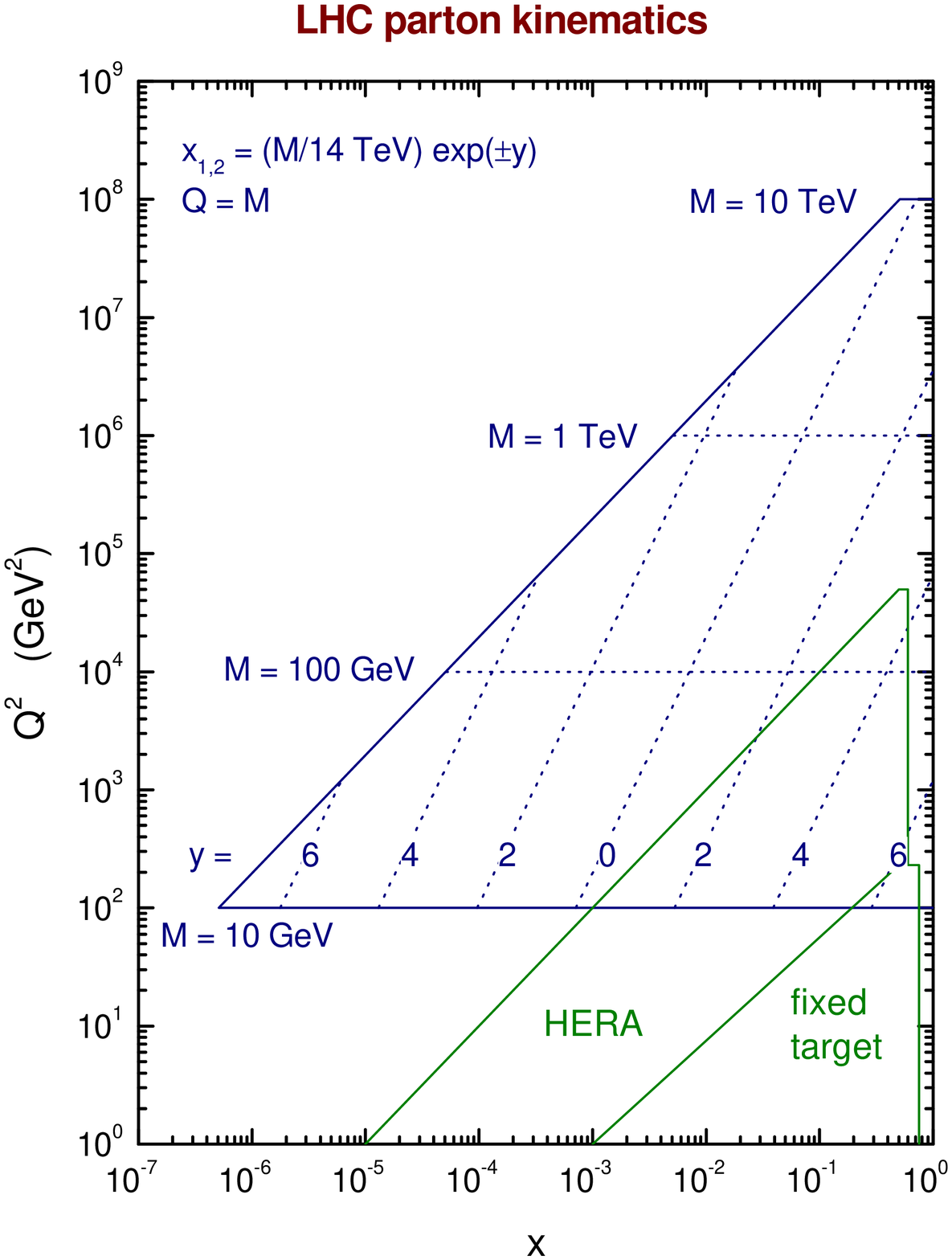}
\end{centering}
%\vspace{-1.0cm}
\caption{Standard Model cross section predictions at hadron-hadron colliders (left), and the parton $x,Q^2$ region probed by the production of a heavy object of mass $M$ and rapidity $y$ at the LHC (right). \label{fig:lhcxq}}
\end{figure}

\section{Parton distributions at the LHC}
\label{sec:LHCimpact}

There are a number of LHC standard-candle processes, $\sigma(W^\pm, Z^0, t\bar t, {\rm jets}, ...)$, that can be used to probe and test pdfs, typically in the range $x \sim 10^{-2 \pm 1},\ Q^2 \sim 10^{4-6}$~GeV$^2$ (see Fig.~\ref{fig:lhcxq}), which is where most New Physics signals (Higgs, SUSY, etc.) are expected.  The total $W$ and $Z$ cross sections provide a particularly important point of comparison between the various pdf sets. A number of factors are relevant, including (i) the rate of evolution from the $Q^2$ of the fitted DIS data to $Q^2 \sim 10^4$~GeV$^2$, driven mainly by $\alpha_S$ and the gluon distribution, and (ii) the mix of quark flavours, since $F_2$ and $\sigma(W,Z)$ probe {\it different} combinations of quark flavours. A very precise measurement of cross section {\it ratios} at LHC (e.g. $\sigma(W^+)/\sigma(W^-)$ and $\sigma(W^\pm)/\sigma(Z)$) will allow these subtle quark flavour effects to be explored.

\begin{figure}[t]
\noindent 
\begin{centering}\includegraphics[width=0.5\columnwidth,keepaspectratio]{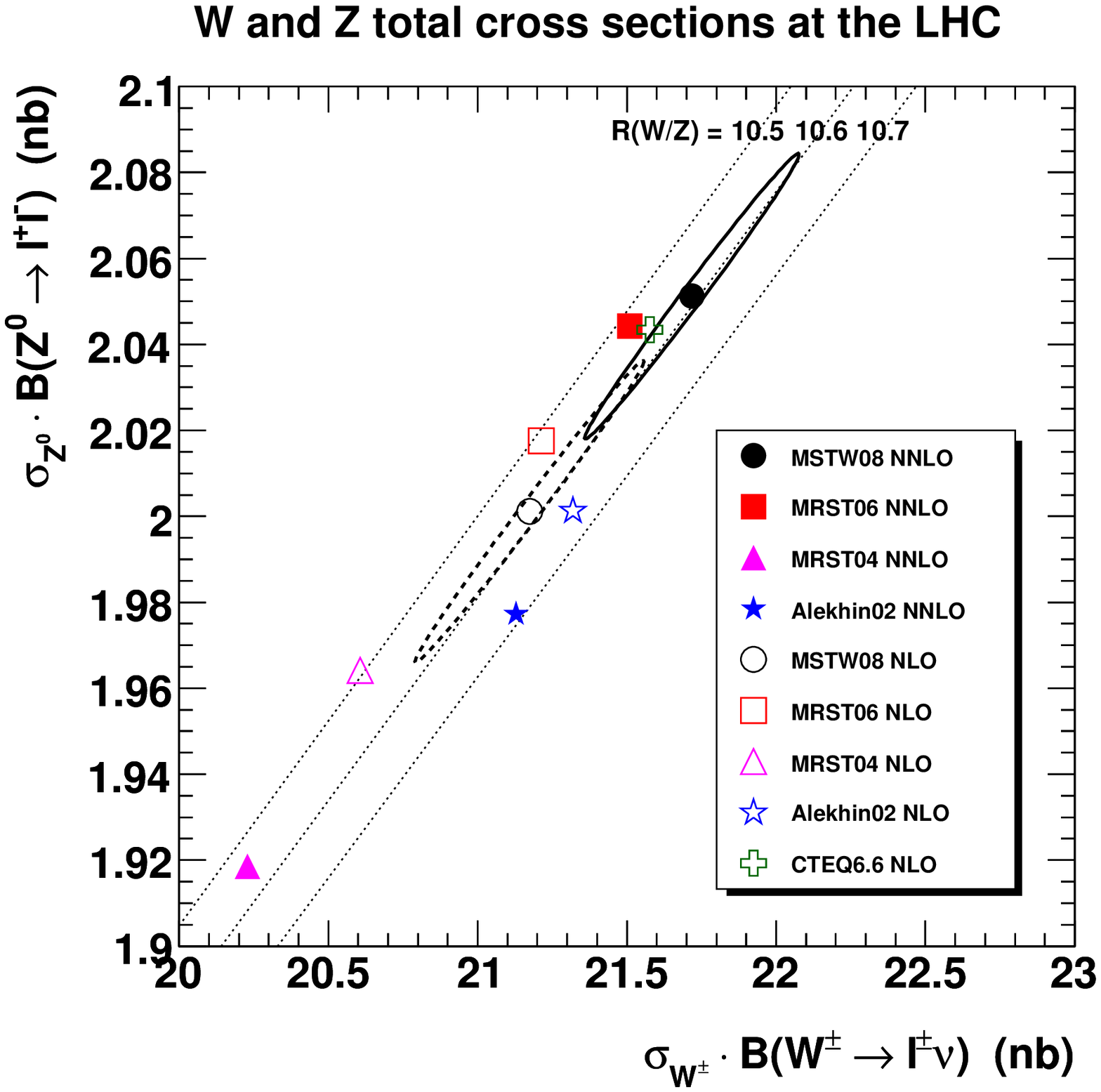}
\includegraphics[width=0.5\columnwidth,keepaspectratio]{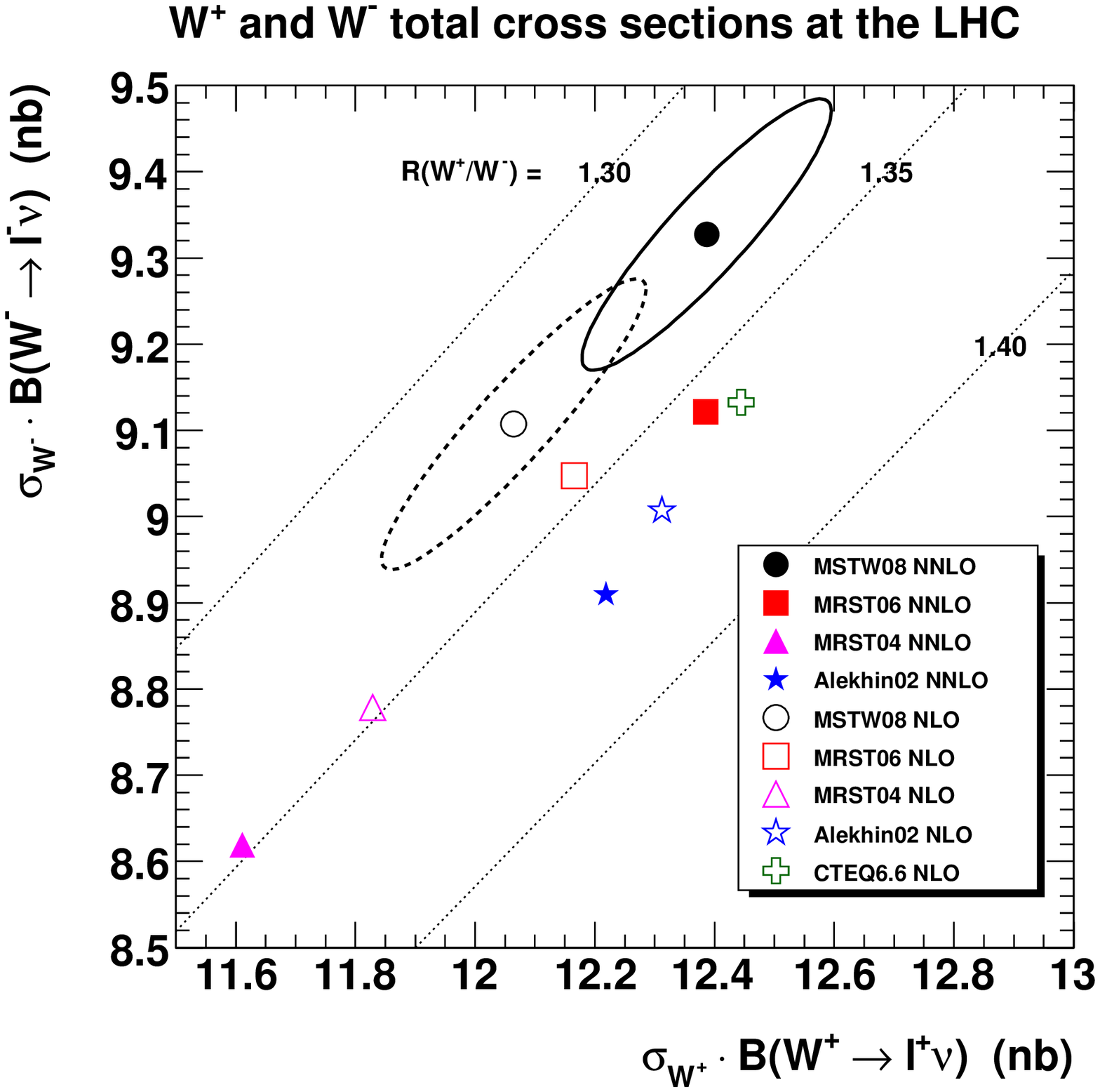}
\end{centering}
\vspace{-0.5cm}
\caption{$W^\pm$ {\it vs.} $Z$ (left) and $W^+$ {\it vs.} $W^-$ (right) total cross sections at the LHC calculated using various past and present pdf sets at NLO and NNLO.  The $1\sigma$ error ellipses are shown for the MSTW 2008 NLO and NNLO pdfs.\label{fig:WZXsecsLHC}}
\end{figure}

By way of example, we show in Fig.~\ref{fig:WZXsecsLHC} a selection of predictions for $\sigma(W^\pm)$ and $\sigma(Z)$ LHC cross sections \cite{MSTW2008}. The error ellipses correspond to the MSTW2008 NLO and NNLO pdf sets. Note that the cross section ratios are determined more precisely than the absolute cross sections themselves. In the case of the $W^+/W^-$ cross section ratio, the overall uncertainty is of order $1\%$, and comes mainly from the uncertainty in the $u/d$ ratio at the relevant $x$  and $Q^2$ values. Note that the change in the cross sections going from MRST2004 to MRST2006 is due to an improvement in the heavy flavour prescription~\cite{Martin:2007bv} discussed earlier, which mainly affects the charm distribution, while the predictions are relatively stable in going from MRST2006 to MSTW2008. The CTEQ6.6 and CTEQ6.5 predictions are very similar, but both are significantly higher than the CTEQ6.1 predictions. Again, this is mainly due to a different treatment of $s,c,b$ quarks in the fit. The CTEQ6.6 LHC predictions are about $+2\%$ higher than MSTW2008, because of slight differences in the quark ($u,d,s,c$) distributions, but overall the predictions agree reasonably well within the quoted uncertainties. Care is however needed in comparing predictions based on different orders of QCD perturbation theory (NLO, NNLO, NNLL-NLO, ...), since the higher-order contributions to the cross sections are not negligible. 

The error ellipses on the MSTW $W$ and $Z$ predictions come from the new `dynamical tolerance' treatment of pdf uncertainties described in \cite{graeme_dis08}. There is an additional uncertainty of the same size from scale variation (quantified in the usual way by varying the scales from $M/2$ to $2M$). Combining these, we predict a total (`$1\sigma$') uncertainty of $\sim\pm 4\%$ on the total $W$ and $Z$ cross sections at LHC, and these could therefore be useful in calibrating the machine luminosity. A more complete discussion of the role of higher-order corrections in cross-section predictions at the LHC can be found in Refs.~\cite{moch,anastasiou}. 

\begin{figure}[th]
%\vspace{-0.4cm}
\centerline{\includegraphics[width=0.46\columnwidth]{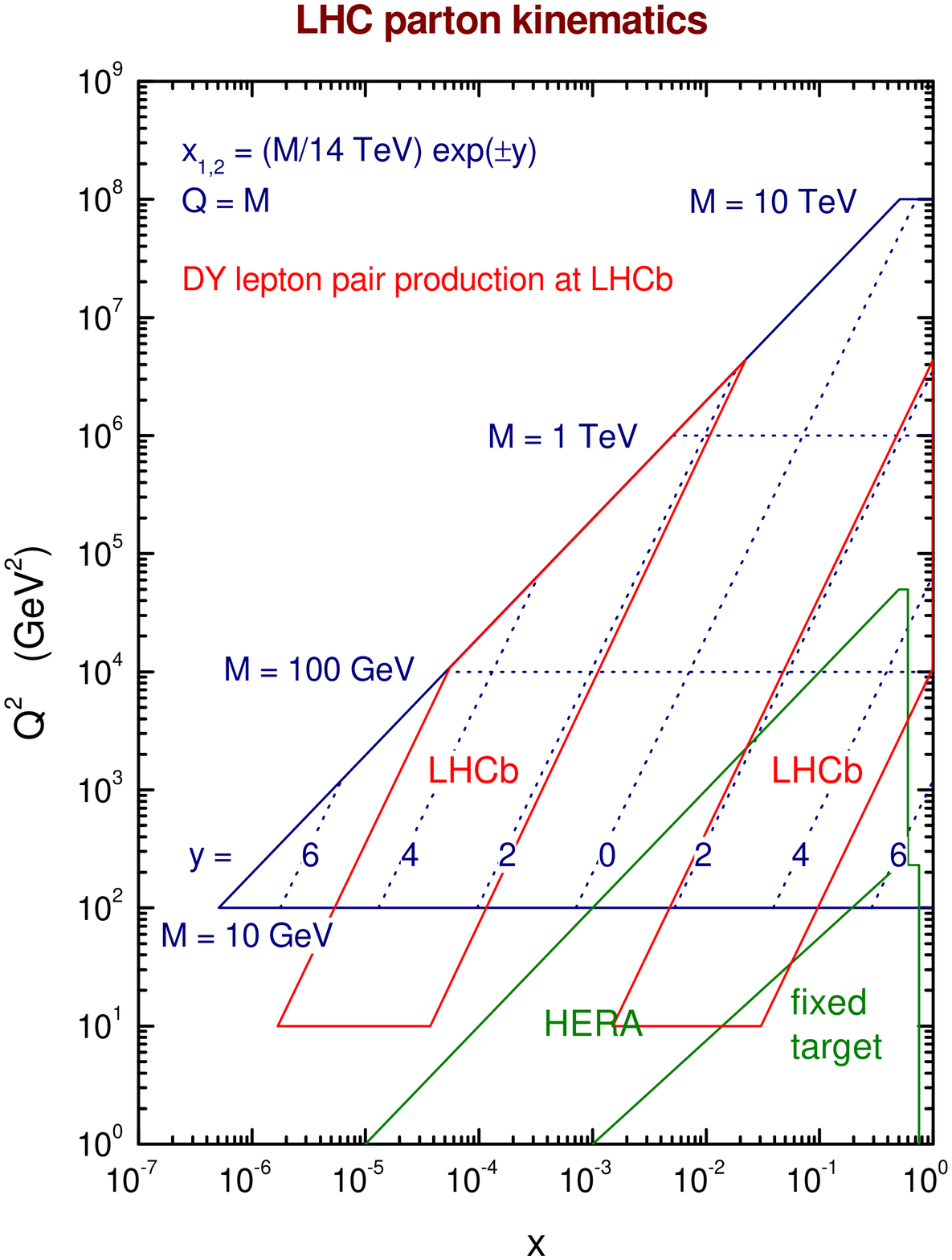}
\hspace{1.0cm}
\includegraphics[width=0.46\columnwidth]{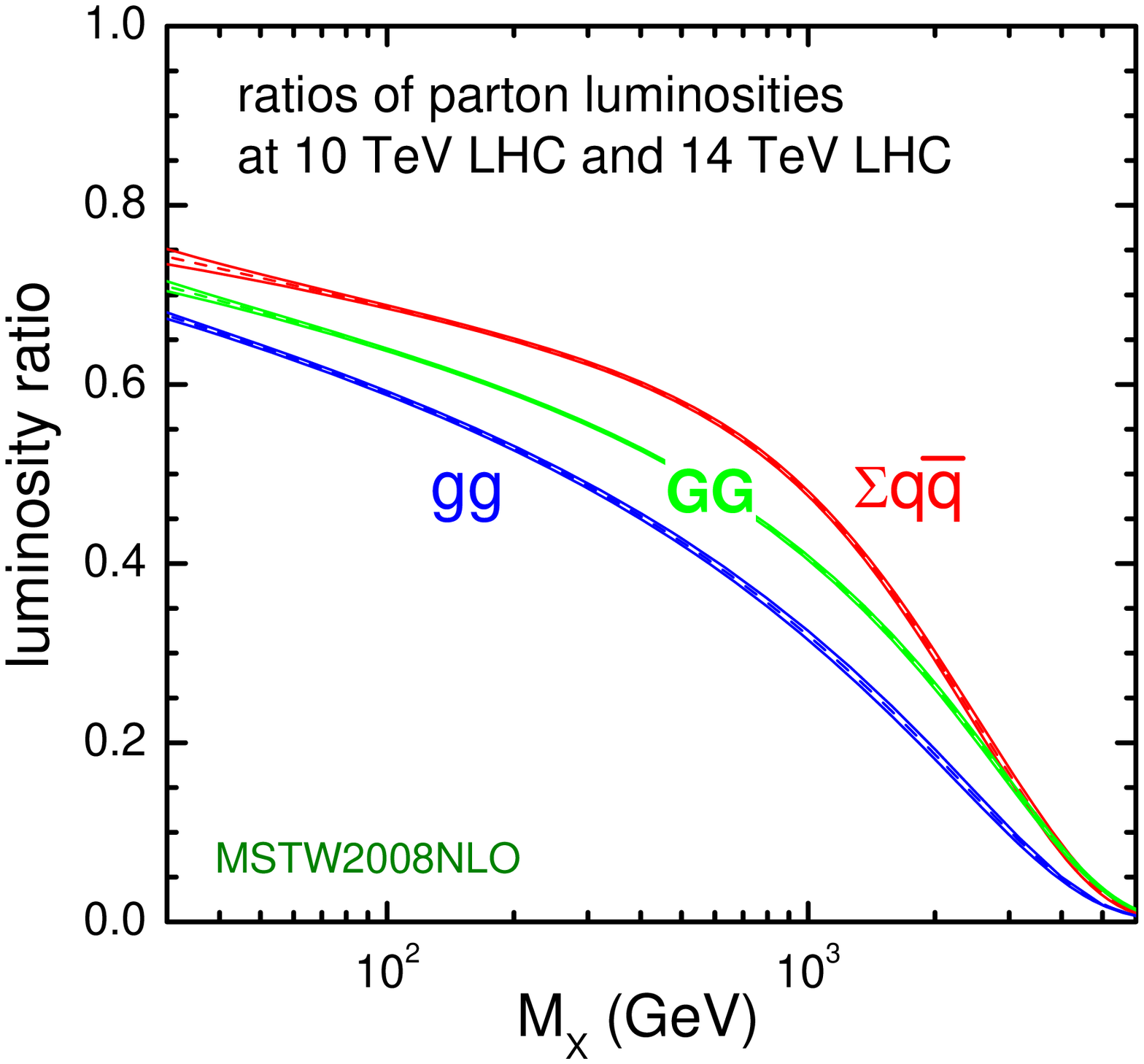}}
\caption{The parton $(x,Q^2)$ region probed by Drell-Yan lepton pair production in LHCb 
(left), and the ratio of $\sum q\bar q$, $gg$ and $GG$ (where $G=g+4/9\sum(q+\bar q)$) parton luminosities at 10~TeV and 14~TeV LHC (right).}\label{fig:lhcbxq}
\end{figure}

It is clear from Fig.~\ref{fig:lhcxq} that in order to probe very small $x$ at the LHC we need to produce relatively light objects at forward rapidity, since then $x \sim (M/\sqrt{s})\exp(-y) \ll 1$. The simplest process to use for this purpose is Drell-Yan (DY) lepton pair production.  At the LHC this requires good detection of low $p_T$ leptons in the forward region. Interestingly, this is precisely the region that will be accessible to LHCb \cite{tara}. Translating the detector acceptance for muon pairs into the $(x,Q^2)$ plane gives the `LHCb' region shown in Fig.~\ref{fig:lhcbxq}. There are two main impacts of such a measurement: (i) quark distributions can be measured in the perturbative domain at smaller $x$ values than at HERA, and (ii) DGLAP evolution over 1--2 orders of magnitude in $Q^2$ can be tested by comparing pdf measurements at the same (small)  $x$ value at HERA and LHCb. Detailed studies are underway to quantify the improvement in pdf precision at small $x$ resulting from the inclusion of such LHCb data in the global fit.

\section{Summary}

In the past few years there has been progress in our understanding of parton distribution functions, and convergence of the various approaches used to determine them. The main distinguishing features of the currently available `precision' pdf sets are (i) how heavy quarks are treated, (ii) how the tolerance for determining pdf error sets is defined, and (iii) whether the Tevatron high$-E_T$ jet data are included in the fit. If they are, then the high-$x$ gluon is slightly larger than the gluon derived from fits which are based on structure function data only. In the context of a {\it full} NNLO global pdf analysis, the NNLO (${\cal O}(\alpha_S^4)$) corrections to the high$-E_T$ jet cross section are still the most important missing ingredient, although their quantitative impact on the current partial-NNLO analysis is not expected to be large. 

The situation regarding the treatment of heavy quark flavour ($c,b$) distributions is now quite satisfactory, with GM-VFNS generally accepted as the correct procedure. Within this framework, there is good agreement with HERA data on $F_2^c$ and $F_2^b$. However, it is important to remember that pQCD-generated heavy flavour distributions may not be the whole story. The issue of additional {\it intrinsic} heavy flavour contributions, dominant at high $x$ where the structure function data are sparse, is still an open question. 

Early data from the LHC will be important for benchmarking a number of Standard Model standard-candle cross sections. In the case of $\sigma(W)$ and $\sigma(Z)$, the (NNL0) cross sections are predicted to approximately $\pm 4\%$ \cite{MSTW2008}. Note that such cross sections are not much smaller at $\sqrt{s} = 10$~TeV energy, since they tend to sample small-$x$ partons that are not changing rapidly with $x$. This is illustrated in the right-hand figure in Fig.~\ref{fig:lhcbxq}, which shows the ratio of the parton luminosities at 10~TeV and 14~TeV for $\sum q \bar q$ (relevant for $W$, $Z$, etc. production), $gg$ (relevant for Higgs, $t \bar t$ etc. production), and $GG$ (with $G = g + 4/9\sum_q(q+\bar q)$, relevant for high$-E_T$ dijet production)  initial states.  

Looking further ahead, a number of LHC measurements have the potential to constrain the pdfs even further. The most interesting appears to be the cross section for relatively low-mass Drell-Yan lepton pairs produced at large rapidity, which may be able to provide information on quark distributions at very small $x \sim 10^{-5} - 10^{-6}$, outside the domain currently accessible at HERA. The LHCb detector appears well suited to this measurement.

\section*{Acknowledgements}
Useful discussions with and input from my MSTW collaborators Alan Martin, Robert Thorne and Graeme Watt are gratefully acknowledged. I would also like to thank the organisers for arranging such an excellent conference.

%------------------------------------------------------------------------------
%       Bibliography
%------------------------------------------------------------------------------
\begin{footnotesize}

\end{footnotesize}
\end{document}